# dPOLY: Deep Learning of Polymer Phases and Phase Transition

Debjyoti Bhattacharya and Tarak K Patra[*]

Department of Chemical Engineering
Indian Institute of Technology Madras, Chennai TN 600036, India

**Abstract**

Machine learning (ML) and artificial intelligence (AI) have the remarkable ability to classify, recognize, and characterize complex patterns and trends in large data sets. Here, we adopt a subclass of machine learning methods viz., deep learnings and develop a general-purpose AI tool - dPOLY for analyzing molecular dynamics trajectory and predicting phases and phase transitions in polymers. An unsupervised deep neural network is used within this framework to map a molecular dynamics trajectory undergoing thermophysical treatment such as cooling, heating, drying, or compression to a lower dimension. A supervised deep neural network is subsequently developed based on the lower dimensional data to characterize the phases and phase transition. As a proof of concept, we employ this framework to study coil to globule transition of a model polymer system. We conduct coarse-grained molecular dynamics simulations to collect molecular dynamics trajectories of a single polymer chain over a wide range of temperatures and use dPOLY framework to predict polymer phases. The dPOLY framework accurately predicts the critical temperatures for the coil to globule transition for a wide range of polymer sizes. This method is generic and can be extended to capture various other phase transitions and dynamical crossovers in polymers and other soft materials.

**Introduction**

Phase transitions in polymers are governed by correlated microscopic interactions and relaxation of their large number of atoms and segments over multiple time and length scales.[1–3] Given the wide variations in microscopic degrees of freedom and macroscopic properties exhibited by polymers, identifying new phases and phase transitions from molecular dynamics trajectories can be challenging. Phase transitions are traditionally characterized via an abrupt change in an order parameter representing a system's local structures. The order parameter is chosen such that it is sensitive to the specific character of a transformation and can provide insightful information about the transition. However, there is no universal choice of order parameter to capture the wide range of phase transformations that take place in polymers and other soft materials. Also, there is no single straightforward approach to identify metastable phases and dynamical crossovers such as vitrification, jamming, gel formation, and localization transition.[3,4] Identifying such crossovers and phase transitions require a priori knowledge of the transformations. Furthermore, there are many materials science problems where no conventional order parameter exists,[5] especially disorder phase transitions such as liquid to liquid, liquid to amorphous solids, and metastable phases are hard to identify. Accurate detection and systematic characterization of phase transitions and dynamical crossovers often require simulations in a very specific ensemble and advanced sampling, which can be computationally expensive. Therefore, discovering new phase transitions and dynamical crossovers requires developing a generic strategy that does not need a priori information. To this front, machine learning methods are powerful in recognizing minute changes in a dataset's trends and patterns. There are many recent emphasis on developing simple yet generic machine learning based strategy for identifying and quantifying phase transition in materials system.[6–10] Motivated by these studies, here we aim to develop a generic machine learning framework for autonomous identification and characterization of phase transition and dynamical crossover in molecular dynamics trajectory of polymers.

There are two approaches for machine learning - unsupervised and supervised methods to derive useful inferences from datasets. In unsupervised machine learning, the learning algorithm receives unlabelled data, wherein the task is to extract features or make a subgroup of data based on specific characteristics (known as clustering analysis). On the other hand, during supervised machine learning, data are supplemented by labels that distinguish a subset of data from the rest of the data set. During supervised machine learning, the learning algorithm learns the data's mapping to its corresponding labels and subsequently predicts the unknown data labels during the prediction cycle. Although many supervised and unsupervised machine learning methods are used to detect phases, phase transitions, and crossovers in condensed matter systems,[11,12] their adaptation in polymers and soft materials are not common. Here, we explicitly focused on a subclass of supervised and unsupervised machine learning tools viz. deep learning for predicting phases in a polymer's molecular dynamics trajectory. We introduce dPOLY, a deep learning framework that captures polymer conformations' variations over a thermophysical process such as cooling, compression, and drying and determines the phases and their crossover point. This framework consists of two major components –

i) an unsupervised deep autoencoder to map the trajectory to a lower dimension, and
ii) a supervised deep neural network (DNN) that is built on the lower-dimensional polymer conformation to predict the phases and phase transition.

The unsupervised deep autoencoder reduces the number of features of a polymer structure by choosing the most important ones that still retain the essence of the polymer structure.[13,14] This reduction of dimensionality of a 3N dimensional polymer conformation improves the DNN classification accuracy. We demonstrate this framework for a prototypical example of a coil to globule transition of a polymer chain undergoing cooling.

Chain conformation in dilute solution, notably the coil−globule transition (CGT), has long been studied due to its theoretical importance and diverse applications.[15] Molecular simulations - Monte Carlo (MC) and molecular dynamics (MD) with either implicit or explicit solvent models and mean-field level theories have been extensively used to understand the influence of chain length, chain stiffness, confinement, and other effects on the CGT transitions, and their order and transition dynamics.[16–27] The CGT is typically determined by an abrupt change in the polymer's macroscopic properties, such as specific heat and radius of gyration. Accurate quantification of CGT transitions requires advanced sampling methods such as integrated temperature sampling (ITS)[16] and Wang-Landau Algorithm.[23] Unlike these conventional approaches, here our goal is to predict the phase transition by analyzing the standard MD simulation trajectory of a polymer without any advanced sampling, to establish and validate a data-driven framework for identifying and characterizing phases and phase transition.

The data are generated by conducting molecular dynamics simulations of a polymer chain in implicit solvent condition by solving its Langevin equation of motion across a wide range of temperatures. An autoencoder neural network is developed using all the polymer conformations that are collecting during the simulated annealing of polymer. Subsequently, a feed-forward back propagation neural network is developed based on the high temperature and low temperature conformation of the polymer. The dPOLY framework accurately predicts coil and globule phases of all the polymer conformations. It also provides an accurate estimation of CGT temperature for a wide range of polymer chain lengths. The crossover temperature exhibits a power law behaviour with the polymer chain length, which agrees with the previous studies. Beyond its current application to the coil to globule transitions in a coarse-grained polymer model, the dPOLY framework is extensible for chemically realistic polymer models with wide variations in molecular structure, chemistry and phases. As in this application, this approach is likely to yield new physical understanding and capturing new phases, including metastable ones for polymers and other soft materials.

**Model and Methods**

The dPOLY framework is developed while testing it to identify and characterize the coil to globule transition of a model polymer system. In this section, we first describe the polymer model and the individual components of the dPOLY framework. Subsequently, we construct the workflow of dPOLY that integrates MD simulation, unsupervised deep learning, and supervised deep learning methods.

**Polymer Model.** We use a generic coarse-grained model of homopolymers for validating the dPOLY framework in identifying coil to globule transition. In this model system, two adjacent coarse-grained monomers of a polymer are connected by the Finitely Extensible Nonlinear Elastic (FENE) potential[28] of the form $V_{FENE} = -\frac{1}{2}KR_0^2\left[1-\left(\frac{r}{R_0}\right)^2\right]$, where, $k = 30\epsilon/\sigma^2$ and $R_0 = 1.5\sigma$. Any two monomers in the system interact via the Lennard-Jones (LJ) potential of the form $V(r_{ij}) = 4\epsilon_{ij}\left[\left(\frac{\sigma}{r_{ij}}\right)^{12} - \left(\frac{\sigma}{r_{ij}}\right)^6\right]$. The $\epsilon_{ij}$ is the interaction energy between any two monomers $i$ and $j$. The size of all the monomers are σ. The LJ interaction is truncated and sifted to zero at a cut-off distance $r_c = 2.5\sigma$ to represent attractive interaction among the monomers.

**Molecular Dynamics Simulations.** Implicit solvent molecular dynamics of polymer chains are simulated for a range of temperatures that incorporate the coil to globule transition. The initial configuration of a polymer chain is placed in a cubic simulation box of fixed dimensions. The simulation box is periodic in all three dimensions. The number density of particles in the simulation box is 0.001. The equation of motion are integrated using the Velocity Verlet algorithm with a timestep of $0.001\tau$, where $\tau = \sigma\sqrt{m/\epsilon}$ is the unit of time. The temperature of the systems is controlled using the Langevin thermostat within the LAMMS simulation environment.[29] The simulation is conducted for polymers of chain length N=30, 50, 100, 200, 300, 400, 500, and 800. We generate polymer trajectories for a range of temperatures starting from T*=4.0 and end at T*=0.5 via successive cooling, equilibration, and production cycles. At each cycle, the temperature is decreased by 0.04T*. Here, the temperature $T^* = Tk_B/\epsilon$ is in reduced LJ unit, where $k_B$ is the Boltzmann constant. For any given temperature, the system is equilibrated for $10^8$ MD steps followed by a production cycle of another $10^8$ MD steps. The data during the production cycles are collected for machine learning and data analytics.

**Unsupervised Deep Autoencoder**. An autoencoder is an unsupervised learning method in which a neural network is used for the task of representation learning. A neural network architecture is designed in such a way that it compresses the correlations in an input dataset to a lower-dimensional representation and consequently decompresses them back to their original correlations. This compression and decompression of data and their correlations are achieved via a bottleneck neural network structure as shown in Figure1. The neural network has two fragments – an encoder and a decoder. The encoder compresses the data into a lower-dimensional latent space representation which is also known as feature space. The decoder expands the lower-dimensional representation into their actual higher dimensional representation. In the current study, the autoencoder compresses 3N positional coordinates of a polymer chain of length N into a lower dimensional latent space representation. The network contains an intermediate "bottleneck" layer that consists of fewer nodes than the input and output layers. This particular architecture forces the network to learn a compact representation of the input data, i.e., the polymers coordinates. Therefore, it allows automatic discovery of polymer configuration features in low dimensional representation and helps classify the phases. Thus, this autoencoder helps visualize and characterize the collective behavior and correlations of a large number of interactive particles in a many-body system in a lower dimension, which is otherwise inconvenient to identify in its actual higher dimension.

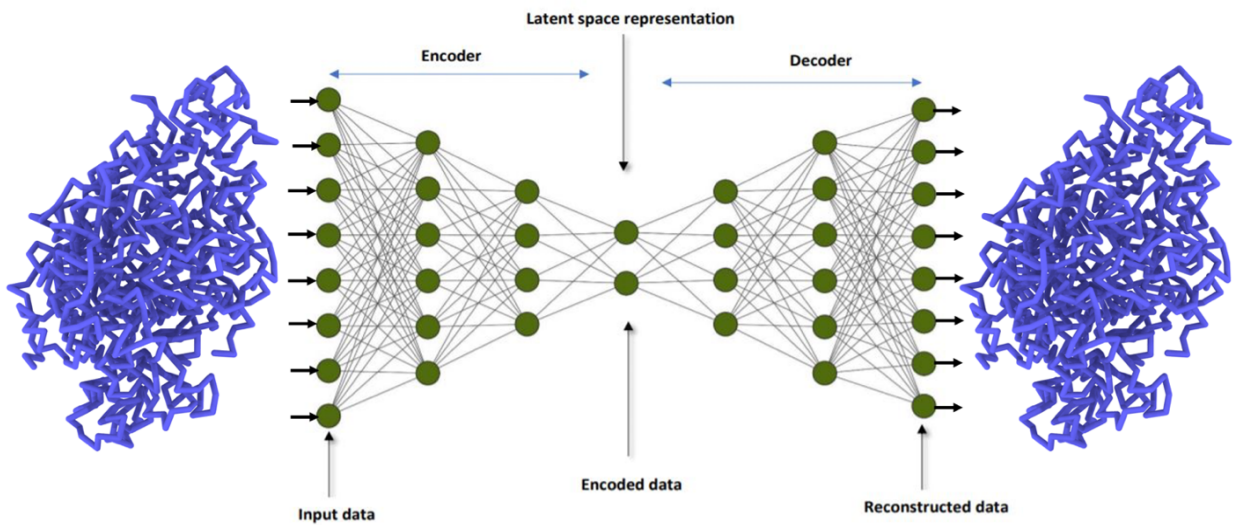

*Figure 1: Autoencoder neural network. It receives the position coordinates of all the monomers of a polymer chain through its input layer neurons. The position coordinates are mapped to a lower dimensional space by the encoder part of the network. The lower dimensional representation is reconstructed to the actual higher dimension i.e., the position coordinate of the monomer of the polymer by the decoder part of the network. The number of nodes in the input layer and the output layer (reconstruction layer) are same. The number of latent variables or the dimension of the feature space depend on the polymer chain length.*

The above task is achieved by a feed-forward back propagation neural network with multiple intermediate layers, as shown schematically in Figure1. An autoencoder of this topology is also known as a deep autoencoder due to the presence of multiple intermediate layers. Within an autoencoder topology, both the encoder and decoder segments have equal number of neurons. Our network structure and the number of neurons in its layers can be written as $3N - n_1 - n_2 - n_3 - n_4 - q - n_4 - n_3 - n_2 - n_1 - 3N$, where each of these indices represents the number of neurons in the corresponding layers. Here, the number of neurons in the input and output layers are *3N* for a polymer of chain length *N*. The input layer and the output layer are identical as they hold the same position coordinates of a polymer chain. The $n_1, n_2, n_3$, and $n_4$ are the numbers of neurons in the four intermediate layers in between the input layer and middle layer. Similarly, $n_4, n_3, n_2$, and $n_1$ are the number of neurons in the four intermediate layers in between the middle layer and the output layer. Therefore, the encoder and decoder part of the network are mirror images of each other. The middle layers has q neurons that represent the dimension of the feature space. The $n_1, n_2, n_3, n_4$, and $q$ are system-specific, and system to system adjustments are done to minimize the error in reproducing the exact structure of a polymer of chain length *N* during the training of the autoencoder.

Each neuron of a given layer is connected to all the neurons of its adjacent layer by weight parameters in a directed acyclic graph. A neuron receives the weighted sum of signals from all the neurons of its previous layer and activate it by an activation function[30] and then feed it to all the neurons in its next layer. We use a rectified linear unit (ReLU) function[31] for activating signals in all the neurons in the network's intermediate layers. No computation is done at the input layer neurons that hold the input values. The output layer neuron activates the signal via a linear activation function and provides the output. The network is trained by a feed-forward backpropagation method. During the training, the weights between all the neurons' pairs are adjusted to minimize the error, which is a measure of the difference between the predicted values and the expected reference values at the output layer neurons. The error is represented by the loss function of the network. We use Adam optimizer,[32] a first order gradient-based optimization method, with a learning rate of 0.001 to optimize the autoencoder's cost function. During the training period, the data is fed in batches to the network repeatedly until a termination criterion is achieved. Typically the training continues until a reasonably low value of the loss function is achieved and it reaches a plateau. We stop the training when the loss function is considerably low and not improving over 4 to 5 cycles. More details of the training and development of a feed-forward backpropagation neural network can be found in our previous works.[33,34] Once the autoencoder is trained, it can be subjected to a polymer conformation for estimating its projection to a lower dimension. The trained autoencoder produces an equivalent polymer trajectory at the output layer and its code values in the middle layer. This code vales in the middle layer representing the lower dimensional mapping of the polymer configurations, which is used for further analysis.

**Supervised Deep Neural Network.** A separate deep neural network is established for the supervised learning of the polymer chain conformations labelled by their state variables. The supervised DNN aims to build a nonlinear heuristic model for the relationship between the input variables (e.g., polymer position coordinates) and output variables (e.g., coil state or globule state). A schematic of a supervised DNN for polymer phase prediction is shown in Figure 2; it consists of neurons organized into an input layer, two hidden layers, and one output layer. The input layer consists of nodes representing polymer coordinates in its feature space as determined by the autoencoder. The output layer contains two nodes representing a pair of state variables that define the phase of a polymer conformation. All the neurons in the intermediate layer activate the signal via the ReLU activation function, similar to the autoencoder. The output layer neurons activate the signal via a sigmoidal activation function[35] and produce the output. We note that a DNN with a large number of parameters often suffers from overfitting. We use a dropout regularization procedure[36] to prevent such overfitting. According to this procedure, the weights of a small fraction of intermediate layer neurons are updated in a given cycle. This prevents neurons from adapting too much. More details of the dropout mechanism can be found elsewhere.[36] We use a commonly used dropout rate within a range of 0.2-0.5. The network is trained by a feed-forward backpropagation method similar to the autoencoder training process. The optimal numbers of neurons in the hidden layers have no universal values but are selected in an effort to maximize the efficiency and accuracy of the DNN. Similar to the autoencoder, Adam optimizer[32] is used with a default learning rate of 0.001 to optimize the cost function of the DNN. The trained DNN can be used to predict the labels, i.e., state variables of a polymer conformation.

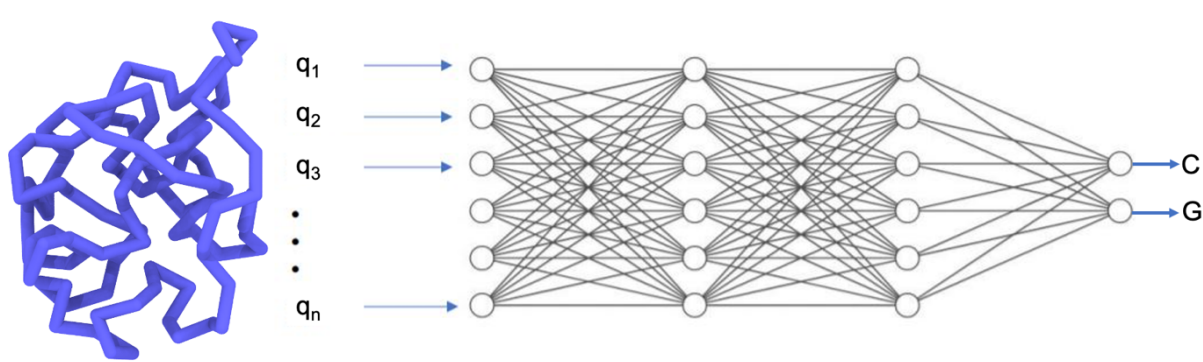

*Figure 2: Supervised deep neural network (DNN). The position coordinates of all the monomers of a polymer conformation in its feature space serve as the input of the neural network. Only the labelled configurations are used for the training of the network. The trained network is used to predict labels of all the frames in a trajectory. The network has two output -C and G that represent the labels i.e., state variables. Here, C=1 and G=0 represent coil phase, while C=0 and G=1 define globule phase.*

**Deep Learning of Polymers (dPOLY).** The workflow of the dPOLY is schematically shown in Figure 3, which consists of five sequential steps as discussed below.

1. We generate a polymer trajectory using MD simulation as described above. The trajectory consists of polymer conformations during the thermophysical process, such as cooling. During the cooling cycle, we collect equilibrium configurations of polymers for 60 different temperatures. At each temperature, 10000 frames are collected.
2. An autoencoder model is built using all the configurations from the MD trajectory for dimensionality reduction and feature selection. The autoencoder is trained by all the frames of the polymer trajectory. Once the autoencoder is trained, it is subjected to all the frames of the same polymer trajectory that consist of configurations for the entire range of temperature. The network produces an equivalent polymer trajectory at the output layer. We extract the signal values from the intermediate 'bottleneck' layer, representing the lower dimensional mapping of the polymer configurations within the temperature range for further study. Here, each data point in the lower dimensional map represents a single polymer configuration within the temperature range.

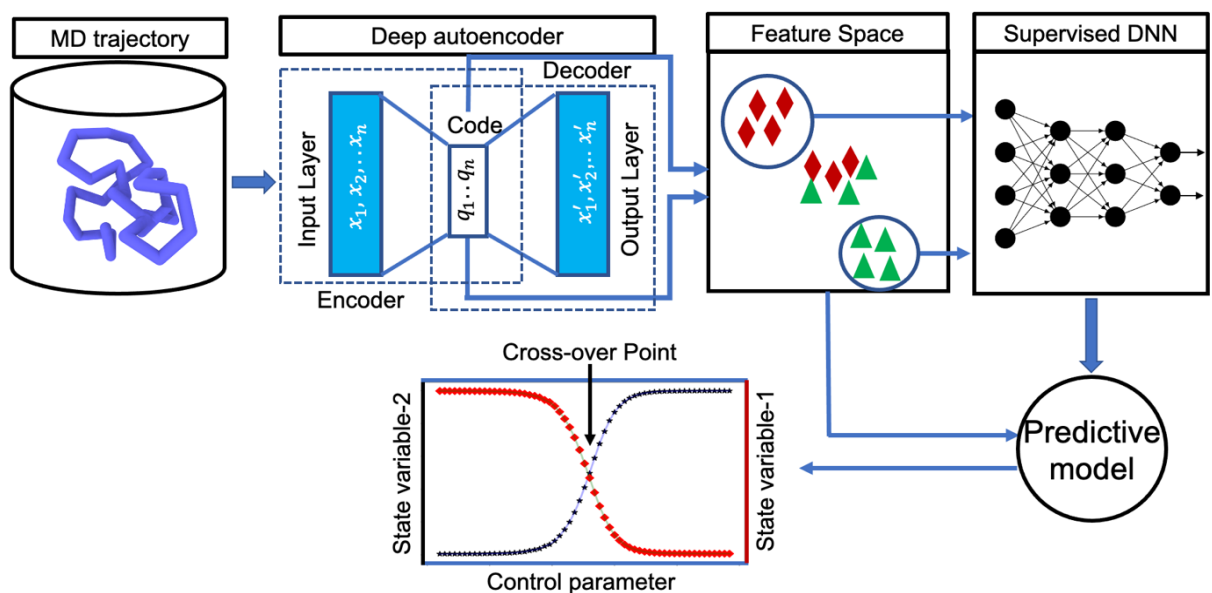

*Figure 3: dPOLY workflow. The workflow consist of five sequential steps – I) Generation of MD trajectory using simulation. II) All the frames of a trajectory is used to build an autoencoder. The trained autoencoder produce code for all the frame in a feature space. III) the frames that are far away from the crossover points are labelled by state variables in the feature space. IV)The labelled frames are then used to developed a DNN predictive model. V) The model predict the labels of all the frames in the feature space, which are further analyzed for accurate identification of crossover points.*

3. We introduce state variables to characterize phases of polymer conformations in the feature space distinctly. We only label the data that are far away from the crossover regions, typically conformations drawn from two extreme temperatures. For the case of one phase transformation, such as coil-globule transition, two state variables *C* and *G* are introduced. The coil state is represented by (C=1 & G=0), and the globule state is represented by (C=0 & G=1) for further machine learning purposes.

4. The labeled data of step-3 is considered as the representation of polymer's pure phases in the molecular dynamics trajectory. The polymer configurations that are labeled in step-3 are then used to develop the DNN. The labeled polymer configurations are randomly selected for the training and testing of the DNN. We use 80 % of labeled data for training and the remaining 20 % to test and validate the network. The training cycles continue until the error in predicting the test data set's state variables are below 2% of their actual values.

5. We use the trained DNN for predicting labels of all the configurations of the MD trajectories that are mapped to the feature space. We note that DNN is trained with selected configurations far away from the crossover points, but it is used to create labels of all the feature space configurations. The labels, i.e., the state variables, are then plotted as a function of the controlling parameter to identify the crossover points. Here, temperature is the controlling parameter for the present case study of coil-globule transition. For each temperature, the state variables of 10000 configurations are predicted by the DNN for averaging purposes. We plot average state variables as a function of temperature. The intersection of these curves is defined as the crossover point of the phase transformation.

The dPOLY, along with its associated autoencoder and DNN, are developed using the Keras deep learning application programming interface (API).[37] The dPOLY is built for a general-purpose AI tool to identify phases, phase transitions, and dynamical crossovers in MD trajectories of polymers undergoing thermophysical treatments such as cooling, drying, and compression.

**Results and Discussion**

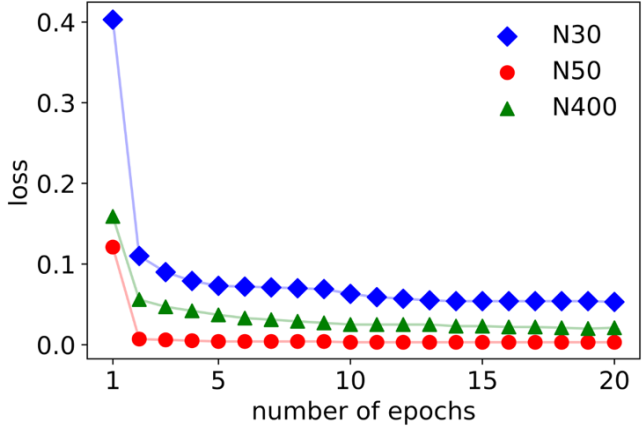

Figure 4: Loss function of the autoencoder during its training for 20 epochs are shown for polymers of chain lengths N =30, 50 and 400.

**Unsupervised learning of polymer structures.** The polymer conformations during the cooling simulations are used to develop an unsupervised deep autoencoder. During the training, 3N position coordinates of a polymer chain are repeatedly fed to the network in batches via the autoencoder's input layer. The output layer neurons of the autoencoder hold the same value as the input layer neurons. The weights between any two neurons are optimized during the training cycle to minimize the loss function of the network, which determine the difference between the input layer and the output layer values of all the 3N position coordinates. We run 20 training cycles for developing the autoencoder model. During the autoencoder training, the loss function is shown in Figure 4 for three representative cases for polymer chain length N =30, 50, and 400. We define loss function as the mean squared error in predicting monomer positions, which can be written as $L_{autoencoder} = \frac{1}{s}\sum_{i=1}^{s}(x_i - x_i')^2$. Here, $s$ is the number of configurations in the training data set. A position coordinate of a monomer in the input layer neuron and output layer neurons is represented as $x_i$ and $x_i'$, respectively. For all the case studies, the loss functions show plateau within the 20 training cycles. This indicates that the autoencoder has learned the polymer structures for the entire range of temperature very accurately. We note that eight autoencoders are developed in this study, each corresponding to a polymer of a specific chain length. As mentioned in the method section, the topology of all the autoencoders is $3N - n_1 - n_2 - n_3 - n_4 - q - n_4 - n_3 - n_2 - n_1 - 3N$, where N is the polymer chain length. The neurons in the intermediate layers $n_1, n_2, n_3$ and $n_4$ are varied from system to system in order to improve accuracy. However, it is possible to fix the values of $n_1$, $n_2$, $n_3$ and $n_4$ across different chain lengths, which will result in a slight variation in loss functions without any significant changes in the featured space representation of the conformations. Similarly, the feature space dimension is also adjusted to improve accuracy.

For example, we identify *q=6* for *N=30* and *q=150* for *N=800* that yields the lowest error in autoencoder's prediction. The exact details of all the autoencoders topologies are summarized in the supplementary information.

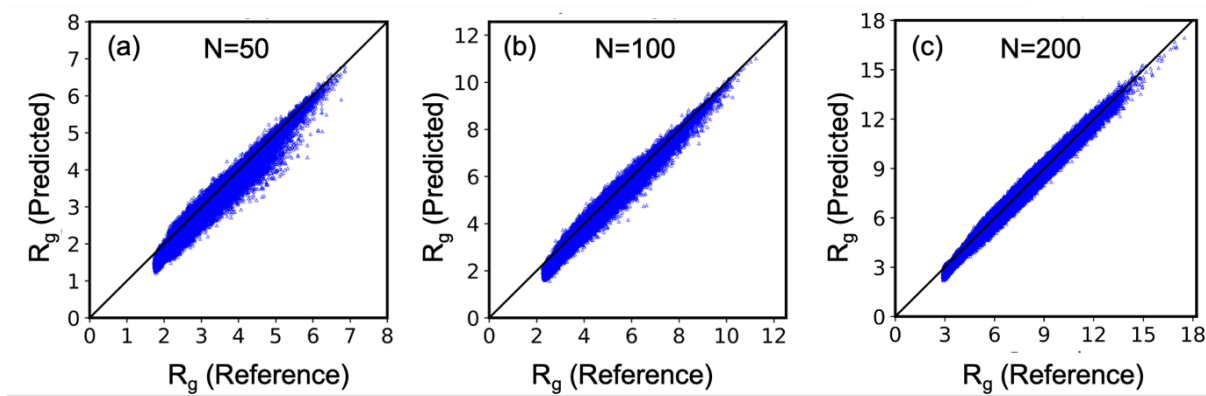

*Figure 5: Performance of autoencoder. Predicted radius of gyration is plotted against the reference values for chain length N =50, 100 and 200 in (a), (b) and (c), respectively. The black lines represent zero mean absolute error in predicted Rg value.*

Now we test the performance of the autoencoder in predicting the structure of the polymer. The snapshot of a randomly chosen polymer conformation fed to the trained autoencoder's input layer and the reconstructed polymer configuration's snapshot in its output layer is shown in Figure 1. It visually suggests that the trained autoencoder replicates a polymer configuration in its output layer with reasonable accuracy. Moreover, we calculate the radius of gyration of individual polymer frames that are passed to the autoencoder along with its predicted configurations. Figure 5 shows the predicted radius of gyration of a polymer conformation as a function of its reference values for three representative cases of *N=50, 100* and *200*. The mean absolute errors for all the cases are within a range of 0.02 to 0.07. Therefore, the predicted Rg values of the autoencoders are in good agreement with the reference values.

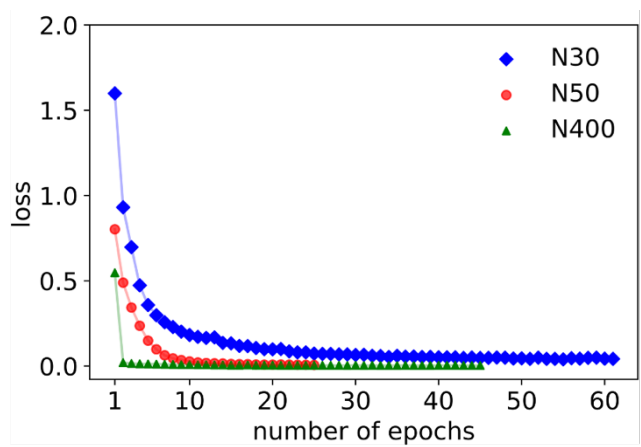

*Figure 6: Loss function during the training cycle of the DNN model for three representative cases of N=30, 50 and 400, respectively.*

**Supervised learning of polymer phases.** We prepare a subset of polymer configurations that are far away from the crossover region, which serves as the training data set for developing a supervised DNN. This subset consists of 20000 configurations of the polymer chains sampled from the highest and lowest temperatures of a cooling cycle. A label - C=1 & G=0 is assigned to the configurations that correspond to the 1st phase (high-temperature configurations), and another label – C=0 & G=1 is assigned for the configurations that correspond to the 2nd phase (low-temperature configurations). We use 50 % of the data for training, and the remaining data are used for testing the performance of the DNN. Figure 6 shows the loss function during the training of the DNN for three representative cases viz., N= 30, 50, and 400. We employ the concept of cross-entropy to define the loss function of the DNN, which is known as the cross-entropy loss function.[38] The cross-entropy loss function of the DNN can be computed as $L_{DNN} = -[t.\ln(p) + (1-t).\ln(1-p)]$ where $t$ is the true value of a label and $p$ is its probability. The probability of a label can be computed as $p(t_i) = e^{t_i}/\sum_{i=1}^{2} e^{t_i}$. Here, $t$ will have a value zero or one for this binary classification problem. It is to be noted that, unlike the state variables' binary value, the cross-entropy function is continuous and differentiable. This particular approach makes it possible to calculate the derivative of the loss function with respect to all the weights in the DNN topology, and thus helps in identifying optimal weights of the network. During the training, the weights and biases of the DNN are updated by using gradient descent via the backpropagation algorithm in order to minimize the cost function. The smaller is the cross-entropy function, the better is the model. We have also adjusted the number of neurons of the intermediate layers of a DNN to incorporate the system to system variation. The exact topologies of the DNNs for different chain lengths are reported in the supplementary information. A perfect DNN model has a cross-entropy loss of zero. Trainings are stopped when the loss function decreases and show a plateau for both test and validation sets. The training cycles vary from 25 to 100, depending on the chain length, as shown in Figure 6. Within the train cycles, the loss function reaches a plateau near zero.

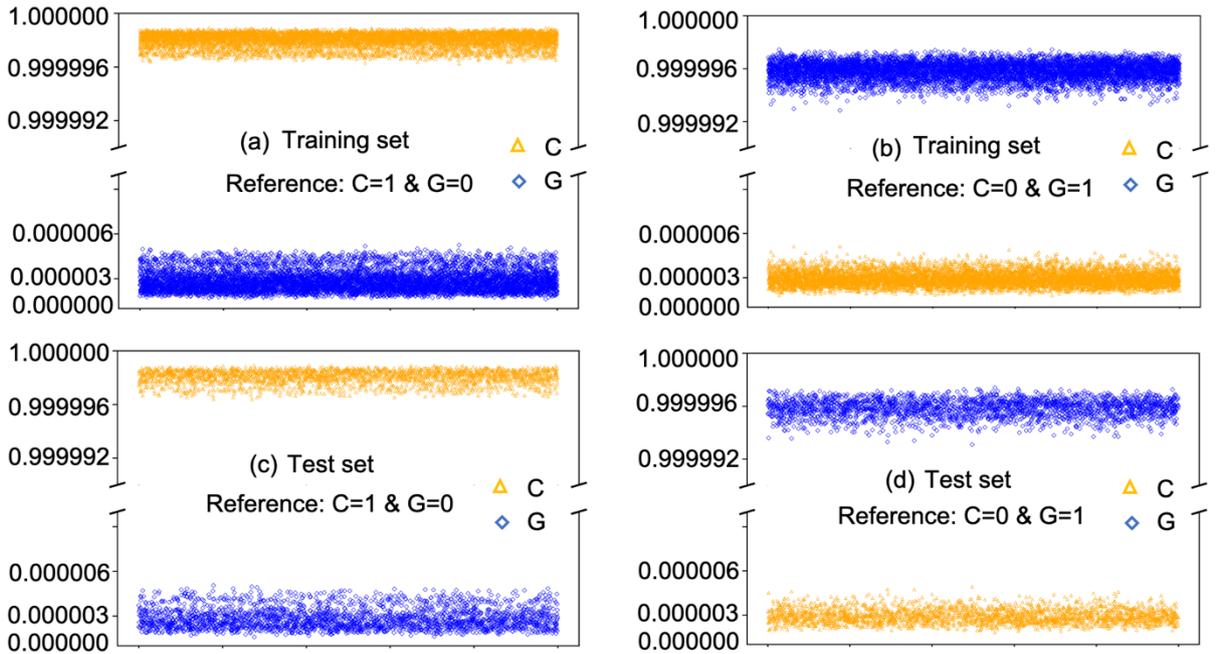

*Figure 7: Performance of the DNN. The output of the DNN - C and G for training and test sets are shown in (a, b) and (c, d) respectively for a representative case of N=400. The references (labels) of conformations are as mentioned in the figures.*

After the completion of training, we test the performance of the DNN for both known and unknown data. The predicted labels of the polymer structures are compared with the reference values for both the training and test sets in Figure 7 for a representative case of a polymer of chain length N=400. For both the training and test sets, we split the data into two subsets, each for a reference phase. For example, the reference values of the labels are C=1 and G=0 for Figure 7a, and DNN predicted labels of all the configurations are C~1 and G~0. For all the cases of polymer chain lengths, the mean absolute error of the DNN prediction is less than 1%. Therefore, we infer that the current DNN model very accurately learns the labels of the polymer structures that represent two distinct phases of the system. This model is further used for predicting the phase transition of polymers during thermal annealing.

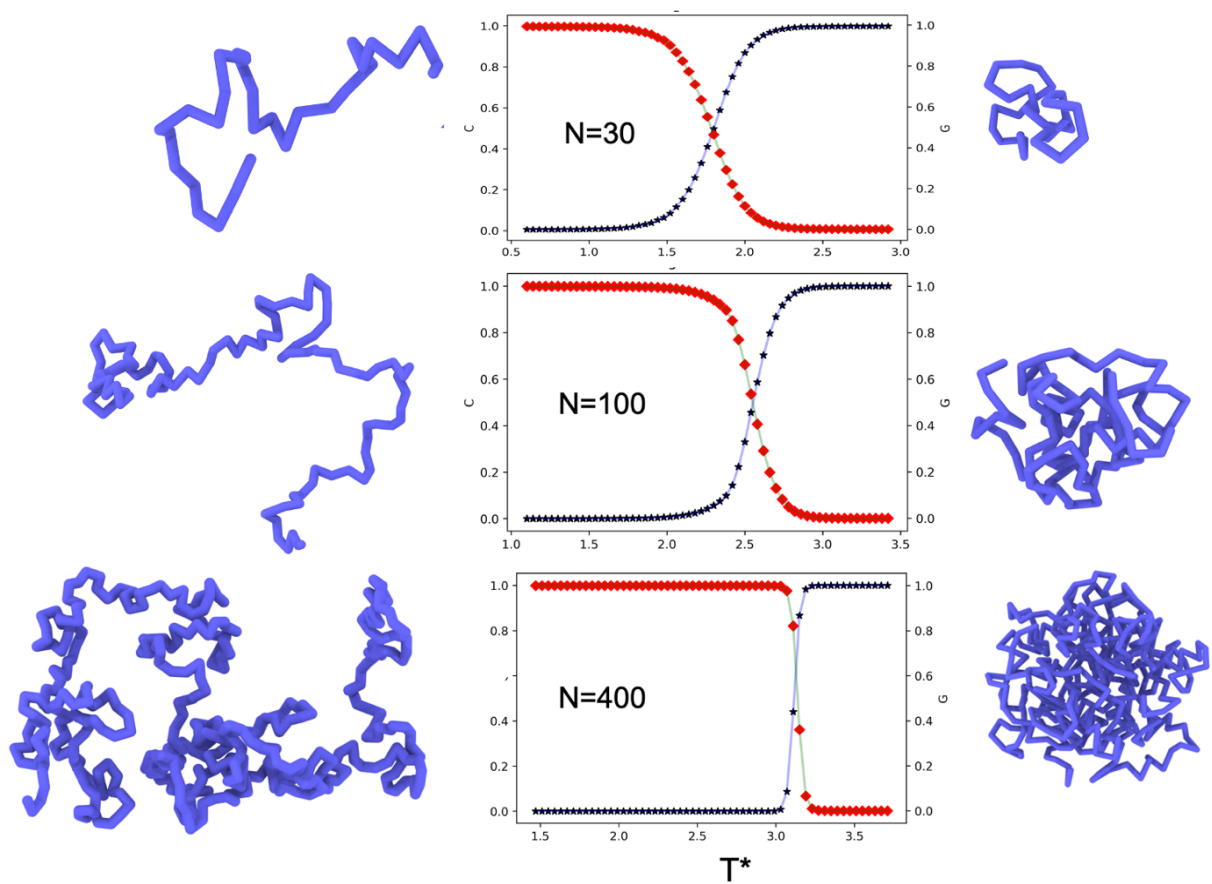

*Figure 8: Machine Learning Predicted Coil to Globule phase transition. The state variable C and G are plotted as a function of temperature for polymer chain length N=30, 100 and 400. Left images correspond to C=1 and G=0. Right images represent C=0 and G=1.*

**Predicting phase transition.** An accurate prediction of phase transition temperature requires well-defined parameters representing a change in the state as a function of temperature. The DNN model can be used for this purpose. Here, we use the trained DNN model to predict the labels of all the polymer configurations across the entire range of temperatures. For each temperature, DNN is used to predict the labels for 10000 configurations. Figure 8 shows the average value of C and G as a function of temperature for three representative cases, N=30, 100, and 200. For all the cases, C and G show a sigmoid type correlation as a function of temperature. At low-temperature range, C remains close to zero, and it increases for an intermediate range of temperature. For high temperature, C plateaus around one. Similarly, G exhibit a value around 1 for the lower temperature range. It decreases in the intermediate range of temperate and shows a plateau near zero for high temperature. We draw continuous curve for both C and G by fitting their respective data using standard spline interpolation scheme. For all the cases, we observe that the C and G curves cross at an intermediate temperature. We infer this intermediate temperature where C and G curves cross each other as the coil to globule transition temperature. The transition temperatures ($T_c$) for all the chain lengths are estimated from the crossover of respective C and G curves and plotted as a function of $1/\sqrt{N}$ in Figure 9. The transition temperature increases with the chain length, and we fit $T_c(N) - T_\theta = a_1/\sqrt{N} + a_2/N$ to further understand the correlation. Here, $a_1$ and $a_2$ are the fitting parameters. The $T_\theta$ is known as the theta temperature that characterizes the crossover of the polymer structure. The solid line in Figure 9 represents the fitted line with $a_1 = -12.44$, $a_2 = 9.47$ and $T_\theta = 3.687$. The phase transition temperature exhibits a primary scaling dependence on the chain length N as $T_c \sim N^{-1/2}$. This prediction is well in agreement with the previously reported correlation derived using traditional approaches such as Wang-Landau and integrated temperature sampling (ITS) methods.[16,23] This verifies that the dPOLY is an effective method for predicting phase transitions in polymers.

We note that a supervised DNN model can be developed based on the 3N position coordinates of a polymer structure without mapping it to a low dimensional feature space via an autoencoder mapping. However, such a DNN based on 3N input layer neurons (3N-DNN) lead to larger error in Tc prediction, and subsequent scaling relation, as shown in the supplementary information. We find that an initial mapping of polymer confirmation to a lower dimensional representation helps improve the efficiency and predictability of a DNN classifier.

## Conclusions

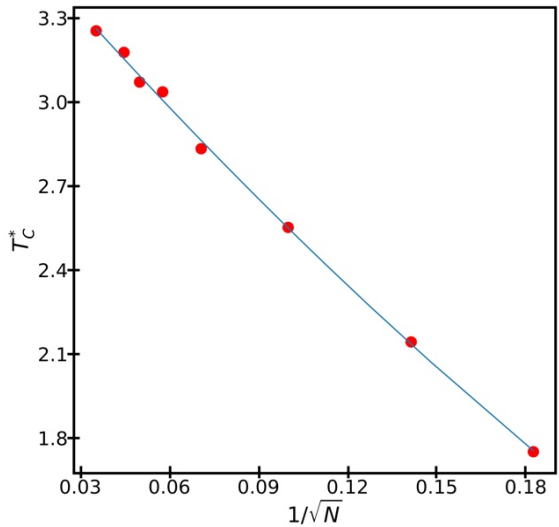

Figure 9: The coil to globule cross over temperature is plotted as a function of $1/\sqrt{N}$, where N is the polymer chain length.

We report an order parameter free approach to identify phases and phase transition in polymers. This approach is based on the deep learning of polymer conformation during a thermophysical process. The dPOLY framework consists of dimensionality reduction and classification of polymer structures across a controlling thermophysical parameter. Within the dPOLY workflow, an unsupervised autoencoder compresses the three dimensional structure of a polymer chain to a lower-dimensional latent space during encoding. During decoding, the model successfully reconstruct the polymer conformation from the lower-dimensional latent space representation to its actual three dimension very accurately. This suggests that the model can generate a unique lower-dimensional trajectory of the 3N dimensional MD trajectory. The lower-dimensional data are then utilized to build a DNN classifier. The DNN classifier model predicts the state variables of all the polymer structures generated during the thermal annealing and thus identify the phase of an unknow structure. The crossover of the state variables as a function of the controlling parameter is estimated as the critical point of the phase transition. We employ dPOLY method for predicting coil and globule phases of a model polymer chain undergoing thermal annealing. The dPOLY framework very accurate predicts the coil to globule transition temperature for a wide range of polymer chain length. The direct use of monomer coordinates as input into the dPOLY underlies the robustness and simplicity of this approach. Although dPOLY is tested for the coil to globule transition, we expect this approach to be able to identify other phase transitions and dynamical crossovers. Especially, the dPOLY is extensible for systems with more than two phases and multiple controlling parameters. Moreover, this deep learning framework does not require a priori knowledge of phase transitions and crossovers and, thus, will be useful for capturing and characterizing complex processes like glass formation and macromolecular crowding, and discovering unknow phases and phase transitions in materials systems. We expect that the present work will lead to the development of more AI tools that are simple yet generic for characterizing structure, dynamic and phase behaviour of polymers and other soft materials.

**Acknowledgment**

The work is made possible by financial support from SERB, DST, Gov of India through a start-up research grant. TKP acknowledges ICSR, IIT Madras for initiation and seed research grants. This research used resources of the Argonne Leadership Computing Facility, which is a DOE Office of Science User Facility supported under Contract DE-AC02-06CH11357. We also used computational facility of the Center for Nanoscience Materials. Use of the Center for Nanoscale Materials, an Office of Science user facility, was supported by the U.S. Department of Energy, Office of Science, Office of Basic Energy Sciences, under Contract No. DE-AC02-06CH11357.